\documentclass[11pt]{article}

\usepackage{epsfig}
\usepackage{amsmath}

\def\ro{\hat\rho_s}

\def\odk #1{(\ref{#1})}
\def\reff #1{Eq.~(\ref{#1})}
\def\refff #1{Eqs.~(\ref{#1})}
\def\ket#1{|#1\rangle}
\def\rr{\vec r}
\def\ra{\rho_1}
\def\rb{\rho_2}

\begin{document}

\begin{center}
{\Large\bf Correlation Functions and Spin} \\[3mm]
{\large Tom\'a\v{s} Tyc}\\[2mm]
{\large \em Dept. of Theor. Physics, Masaryk University,  
         611 37 Brno, Czech Republic} \\
         email: tomtyc@physics.muni.cz\\[2mm]
June 2000\\[1mm]
\end{center}
{\bf Abstract:} The $k$-electron correlation function of a free chaotic electron beam is derived
with the spin degree of freedom taken into account. It is shown that it can be
expressed with the help of correlation functions for a polarized electron beam
of all orders up to $k$ and the degree of spin polarization. The form of the
correlation function suggests that if the electron beam is not highly
polarized, observing multi-particle correlations should be difficult.
The result can be applied also to chaotic photon beams, the degree of spin
polarization being replaced by the degree of polarization.\\[7mm]

\section{Introduction}

Although the theory of multi-particle correlations in free-electron beams
is relatively well developed 
(see e.g.~\cite{silverman1,silverman2,endo,saito,ban1,tyc}),
up to now there has not been much attention devoted to
the electron spin. We feel this as a deficiency because the spin degree of
freedom is a significant feature of the electron and should be therefore taken
into account. In this paper we try to reduce this deficiency for one
particular case that is, as we believe, the most typical one in electron
correlation experiments. Namely, we calculate the multi-electron correlation
function for an electron chaotic beam with an arbitrary degree of spin
polarization. 

Moreover, the presented theory can be applied not only to chaotic electron
(or generally spin-1/2-particle) beams, but also to a chaotic photon
beams because the corresponding Hilbert space of possible photon
polarization states is two-dimensional precisely as is the Hilbert space of the
spin-1/2 states. We will speak about electrons for brevity, but
the argumentation and the results can be applied also to the case of photons.

\section{Chaotic state}

The reason why we concentrate on the chaotic state is that we believe it
is a good approximation for a state produced by the most coherent electron
source available nowadays -- the field-emission tip.
This source is the only candidate for electron correlation experiments at the
present time \cite{silverman2} because it offers both high brightness and
a relatively 
monochromatic energy spectrum. The electrons emitted from a field-emission
tip originate from a quasi-equilibrium state in the metal very close to
a thermal state, which is an example of the chaotic state. We do not expect
an additional coherence to come into existence during the tunneling process
and therefore suppose that the state remains chaotic also outside the
metal\footnote{this might be no longer true for a polarized field-emission
source \cite{europium}, where a correlation between the spin and coordinate
can come into existence; however, it can be shown that our results remain valid
as long as the energy spectra of the spin components ``up'' and ``down''
are close to each other}.

The chaotic state has been first introduced by Glauber for a mode of the
electromagnetic field \cite{glauber} as a state 
with a maximum entropy for a given occupation number of this mode.
We generalize this definition to an electron field with the spin taken into
account: the chaotic state is a state of a maximum entropy for given
occupation numbers of the individual modes of the field and for a given spin
polarization.

From the condition of the maximum entropy
it follows that there is no correlation between the coordinate and
the spin components of the chaotic state, which allows to treat
the spin degree of freedom separately from the coordinate degrees of freedom.
In calculating the correlation functions one can avoid in this way the formalism
of spin-dependent electron field operators and use a relatively simple
argumentation based on the probability theory. As will be seen in the
following, this enables to express the correlation functions
of a partially spin-polarized chaotic electron beam in terms of 
the degree of spin polarization and the correlation functions corresponding to
a completely polarized beam, the form of which is known \cite{saito,tyc}.

For our purpose it is fully sufficient to define the $k$-electron correlation
function $O^{(k)}(\rr_1,t_1,\dots,\rr_k,t_k)$ as the probability of detecting
$k$ electrons at the $k$ space-time points $(\rr_1,t_1)$, $(\rr_2,t_2)$, \dots,
$(\rr_k,t_k)$. To get a more compact form of the equations, the shortened
notation $O^{(k)}_{12\dots k}$ instead of $O^{(k)}(\rr_1,t_1,\dots,\rr_k,t_k)$
will be used, each index $i$ standing for one space-time point $(\rr_i,t_i)$.
Analogous correlation functions for a spin-polarized beam will be denoted
by $G$ instead of $O$, so for example the two-electron correlation function
for a polarized beam is $G^{(2)}_{1,2}$.
For a more precise definition of correlation functions in terms of electron
field operators see e.g.~\cite{silverman1,tyc}.

\section{Two spin-polarized sources}\label{III}

Consider an electron that has been emitted from the source.
Its state can be described by the spin density operator $\ro$
that is represented by a Hermitian matrix of the second order
in any orthonormal basis. As $\ro$ can be diagonalized by a
unitary transformation, there exists an orthonormal basis
$\{\ket 1,\ket 2\}$ in which $\ro$ has the diagonal form
\begin{equation}
  \ro=\left(\begin{array}{cc}\rho_1&0\\0&\rho_2\end{array}\right)
   =\rho_1 |1\rangle\langle 1|+\rho_2 |2\rangle\langle 2|
\label{spindiag}\end{equation}
The states $\ket 1$, $\ket 2$ correspond to the spin orientation
``up'' and ``down'', respectively, with respect to some particular
axis $\vec a$ in space. 
In the case of photons, the states $\ket 1$, $\ket 2$ would describe
two orthogonal polarizations, e.g.~two linearly polarized waves with
the polarization planes perpendicular to each other or a pair of the left-
and right circularly polarized waves, depending on the properties of the
source.

It is useful to express the probabilities $\rho_1$ and $\rho_2$ in terms of the
degree of polarization. This quantity is defined as
$P=(\rho_1-\rho_2)/(\rho_1+\rho_2)=\rho_1-\rho_2$ (provided that
$\rho_1\ge\rho_2$), which yields
\begin{equation}
  \rho_1=\frac{1+P}2,\quad\rho_2=\frac{1-P}2.
\label{rho}\end{equation}

Now, the spin state of the ensemble of electrons coming from the source
is completely described by the density operator \odk{spindiag}.
As can be seen from its form, $\ro$  corresponds to the situation as if
just two types of electrons were emitted from the source: first, electrons
polarized up with respect to the axis $\vec a$ and second, electrons
polarized down with respect to $\vec a$. The probabilities that an electron
emitted from the source is of the first or second type are $\rho_1$ or $\rho_2$,
respectively. The~fact that the spin density operator $\ro$ is diagonal means
that there is no correlation between the two spin components up and down. At
the same time, there is no correlation between the spin and the coordinate
because the complete multi-electron state is chaotic. This allows the following
consideration to be made.

We substitute formally the original electron source $S$ by two independent
sources $S_{\rm up}$ and $S_{\rm down}$ that emit electrons
polarized up and down with respect to the axis $\vec a$, respectively.
If this should be correct, the sources $S_{\rm up}$ and $S_{\rm down}$ must
have the same properties (of course except for the spin) as the original
source $S$ has.  This implies for example that they are located at the~place
of the~original source
$S$ and have the same energy spectrum. Moreover, the emission intensities of
the sources $S_{\rm up}$ and $S_{\rm down}$ must be equal to $\rho_1$ and
$\rho_2$ times the~intensity of the~original source $S$, respectively.
In the following, the idea of substituting formally the original source by two
polarized ones will be used for a direct derivation of the correlation function
for a partially polarized electron beam. To see the idea more clearly,
we will consider the simple case $k=2$ first and then go over to a general
$k$.

\section{Two-electron correlation function}\label{2-el}

The~two-electron correlation function expresses the~probability of the~event of
detecting two electrons at the~ space-time points $(\rr_1,t_1)$ and
$(\rr_2,t_2)$.
This event can happen in one of four ways that are distinguishable in
principle because the~coordinate and spin operators mutually commute:
\begin{enumerate}
\item  the~spins of both electrons are oriented up with respect to the axis
       $\vec a$,
\item  the~spins of both electrons are oriented down,
\item  the~spin of the~electron at $(\rr_1,t_1)$ is oriented down,
the~spin of the~electron at $(\rr_2,t_2)$ is oriented up,
\item  the~spin of the~electron at $(\rr_1,t_1)$ is oriented up,
the~spin of the~electron at $(\rr_2,t_2)$ is oriented down.
\end{enumerate}
If cases 1 or 2 occur, then according to the previous section we
deal with two electrons from the same polarized source. Therefore
the two-electron correlation function is equal to the~analogous correlation
function $G_{12}^{(2)}$ for polarized electrons. On the~other hand, if 
cases 3 or 4 occur, we deal with two electrons from two independent,
oppositely polarized sources. The~electrons are then completely
uncorrelated and the~correlation function is equal to the~product of
one-electron correlation functions, i.e., $G^{(1)}_{1}G^{(1)}_{2}$.
As the~probability that one electron is polarized up or down is $\rho_1$
or $\rho_2$, respectively, the~probabilities of the~cases 1.--4. are
$\rho_1^2$, $\rho_2^2$, $\rho_1\rho_2$, and $\rho_1\rho_2$, respectively.
The~total correlation function $O^{(2)}_{12}$ can be then written as
the~weighed average of the polarized correlation functions,
\begin{equation}
O^{(2)}_{1,2}
    =(\rho_1^2+\rho_2^2)\,G^{(2)}_{1,2}+ 2\rho_1\rho_2\,G^{(1)}_{1}G^{(1)}_{2},
\label{O2}\end{equation}
which is the desired result for $k=2$.

\section{$k$-electron correlation function}\label{V}

In the derivation of the $k$-electron correlation function for
a partially polarized chaotic electron beam we proceed in a 
completely analogous way. If $k$ electrons at the
space-time points $(\rr_1,t_1),\dots,(\rr_k,t_k)$  should be detected,
there are $2^k$ possibilities how they can be polarized (instead of the four
possibilities discussed in the previous section). 
We denote each of them by the~sequence $s_1,s_2,\dots,s_k$,
every $s_i$ expressing the~spin polarization of the~electron at the~point
$(\rr_i,t_i)$ and having one of two possible values, 1 for spin up and or 2 for
spin down. The~probability $P(s_1,\dots,s_k)$ that the~electrons have
polarizations $s_1,\dots,s_k$ is equal to $\ra^{n_1}\rb^{n_2}$, where $n_1$
and $n_2$ expresses how many times there appears 1 and 2 among the
numbers $s_1,\dots,s_k$, respectively. 
If $O_{1,\dots,k}^{(k)}(s_1,\dots,s_k)$ denotes
the~$k$-electron correlation function for this particular spin combination,
the~total $k$-electron correlation function can be written as
\begin{equation}
  O_{1,\dots,k}^{(k)}=\sum_{s_1,\dots,s_k}
  P(s_1,\dots,s_k)\,O_{1,\dots,k}^{(k)}(s_1,\dots,s_k),
\label{rozklad}\end{equation}
the~sum being made over all the~possibilities $s_1,\dots,s_k$.
Now, if the~spin polarizations of the~electrons are
$s_1,\dots,s_k$, the~situation is the~same as if we dealt
with two independent sets of electrons --- one
set of $n_1$ up-polarized electrons originating from the source $S_{\rm up}$
and another set of $n_2$ down-polarized electrons originating from the
source $S_{\rm down}$. The~correlation function
$O_{1,\dots,k}^{(k)}(s_1,\dots,s_k)$ factorizes
therefore into a~product of two correlation functions for polarized electrons:
\begin{equation}
  O_{1,\dots,k}^{(k)}(s_1,\dots,s_k)
  =G^{(n_1)}\left(\{\rr,t\}_{\rm up}\right)\,
   G^{(n_2)}\left(\{\rr,t\}_{\rm down}\right).
\label{factorize}\end{equation} 
Here $\{\rr,t\}_{\rm up}$ and $\{\rr,t\}_{\rm down}$ denote the
sets of points at which the~electrons are polarized up and down, respectively.
Substituting \reff{factorize} into \reff{rozklad} and re-arranging the sum,
we can write the correlation function $O^{(k)}_{1,\dots k}$ as follows:
\begin{multline}
  O^{(k)}_{1,\dots,k}=\left(\ra^k+\rb^k\right)G_{1,\dots,k}^{(k)}\\
  +\left(\ra^{k-1}\rb+\rb^{k-1}\ra\right)
  \left(G_{2,\dots,k}^{(k-1)}G_{1}^{(1)}
  +G_{1,3,\dots,k}^{(k-1)}G_{2}^{(1)}+\dots
  +G_{1,\dots,k-1}^{(k-1)}G_{k}^{(1)}\right) \\
  +(\ra^{k-2}\rb^2+\rb^{k-2}\ra^2)
  \left(G_{3,\dots,k}^{(k-2)}G_{1,2}^{(2)}
  +G_{2,4,\dots,k}^{(k-2)}G_{1,3}^{(2)}\dots
  +G_{1,\dots,k-2}^{(k-2)}G_{k-1,k}^{(2)}\right)+\dots
\label{k}\end{multline}
To see the~structure of such a~series better, we write down the~three- and
four-electron correlation functions for illustration:
\begin{equation}
  O^{(3)}_{1,2,3}=\left(\ra^3+\rb^3\right)G_{1,2,3}^{(3)}
  +\left(\ra^{2}\rb+\rb^{2}\ra\right)
  \left(G_{1,2}^{(2)}G_{3}^{(1)}
  +G_{1,3}^{(2)}G_{2}^{(1)}
  +G_{2,3}^{(2)}G_{1}^{(1)}\right),
\end{equation}
\begin{multline}
  O^{(4)}_{1,2,3,4}=\left(\ra^4+\rb^4\right)G_{1,2,3,4}^{(4)}\\
  +\left(\ra^3\rb+\rb^3\ra\right)
  \left(G_{1,2,3}^{(3)}G_{4}^{(1)}
  +G_{1,2,4}^{(3)}G_{3}^{(1)}
  +G_{1,3,4}^{(3)}G_{2}^{(1)}  
  +G_{2,3,4}^{(3)}G_{1}^{(1)}\right)\\
  +2\ra^2\rb^2
  \left(G_{1,2}^{(2)}G_{3,4}^{(2)}
  +G_{1,3}^{(2)}G_{2,4}^{(2)}
  +G_{1,4}^{(2)}G_{2,3}^{(2)}\right).
\end{multline}
In this way the~$k$-electron correlation function for partially
polarized electrons is expressed in terms of the~one-, two- etc. up to the
$k$-electron correlation functions for polarized electrons and the~degree of
polarization [that is connected with $\ra$, $\rb$ via the~relations \odk{rho}].

The considerations made in the sections \ref{III}. -- \ref{V}. as well as the
results \odk{O2} and \odk{k} can be applied step by step also to a chaotic
photon field because up to now we have not supposed anything about the quantum
statistics of the electrons. In fact, this statistics is hidden in the
spin-polarized correlation functions $G$ and in this way it is reflected also
in the correlation function $O^{(k)}_{1,\dots,k}$. As has been mentioned,
the similarity between electrons and photons in this sense comes from
the same dimensions of the photon polarization Hilbert space and the electron
spin Hilbert space.

We return to the electrons again.
According to \cite{saito,tyc}, the $k$-electron correlation function for
a spin-polarized chaotic electron beam can be expressed as
\begin{equation}
G_{1,\dots,k}^{(k)}=G^{(1)}_1G^{(1)}_2\cdots G^{(1)}_k
  \det \hat\gamma
\label{DETT}\end{equation}
where $\hat\gamma=(\gamma_{ij})$ is a matrix composed of the complex degrees
of coherence $\gamma_{ij}$ at the points $(\rr_i,t_i)$ and $(\rr_j,t_j)$.
Combining \refff{k} and \odk{DETT}, we arrive at the explicit form for
the correlation function for a chaotic electron beam with an arbitrary spin
polarization.

\section{Influence of polarization on multi-electron correlations}

To see how the spin polarization influences the correlations in an electron
beam, we first return to the case of $k=2$. According to \reff{DETT}, the
two-electron correlation function for a spin-polarized chaotic electron beam is
equal to
\begin{equation}
G^{(2)}_{1,2}=G^{(1)}_{1}G^{(1)}_{2}(1-|\gamma_{12}|^2)
\end{equation}
where we used the fact that $\gamma_{11}=\gamma_{22}=1$.
With the help of \refff{rho} and \odk{O2}, we then get for $O^{(2)}_{1,2}$
\begin{equation}
   O^{(2)}_{1,2}=G^{(1)}_{1}G^{(1)}_{2}
                \left(1-\frac{1+P^2}{2}|\gamma_{12}|^2\right).
\label{corr22}\end{equation}
If there were no correlation between the detection probabilities at the points
$(\rr_1,t_1)$ and $(\rr_2,t_2)$, the correlation function $O^{(2)}_{1,2}$ would
be simply equal to the product of the one-electron correlation functions
$O^{(1)}_{1}=G^{(1)}_{1}$ and $O^{(1)}_{2}=G^{(1)}_{2}$.
Therefore the second term in the parentheses in \reff{corr22} is responsible
for the two-electron correlation. As we can see, this term increases with
the increasing degree of polarization of the beam, varying between one half
for an unpolarized beam and unity for a completely polarized beam.
For electrons, this result has been known \cite{silverman1} but it has been
derived heuristically only until now.
For photons a similar effect of the polarization on the correlation function
is known \cite{mandel}.

\begin{figure}[htb]
\begin{center}
\mbox{\epsfxsize=9cm\epsfbox{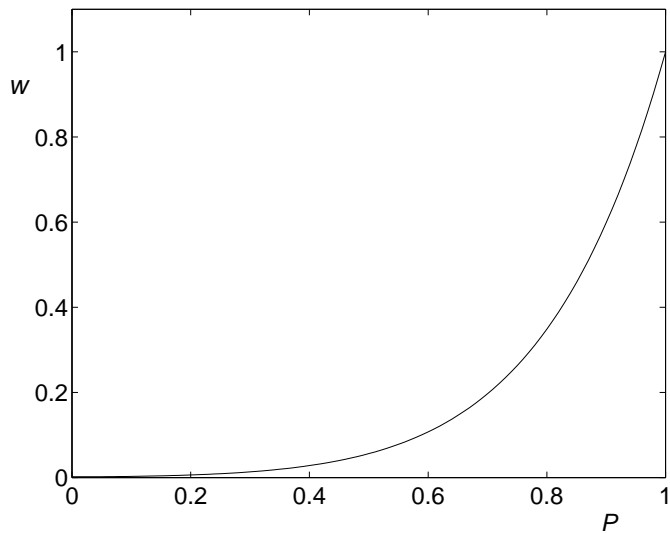}}
\caption{\small The factor $w=\ra^k+\rb^k=[(1+P)/2]^k+[(1-P)/2]^k$ expressing
the intensity of the $k$-electron correlation as a function of the degree of
polarization $P$ for $k=10$. As the figure shows, if $w$ should be comparable
to unity, making thus the $k$-electron correlations observable, one needs a
beam of a relatively high degree of polarization.}
\label{fig}
\end{center}
\end{figure}

Next we go over to the case of an arbitrary $k$. According to \reff{k}, the
factor $w=\left(\ra^k+\rb^k\right)$ expresses the weight of the $k$-electron
correlation in the partially polarized beam compared to a completely polarized
beam because the $k$-electron correlation is given
just by the correlation function $G^{(k)}_{1,2,\dots,k}$. It is evident
that for large $k$ the factor $w$ is small as soon as $P$ differs even slightly
from unity. Figure \ref{fig} shows the dependence $w(P)$ for $k=10$.
For example if $P=0.7$, then $w$ is equal to only about 0.2, so the
ten-electron correlation is reduced to one fifth with respect to a polarized
beam. Thus we must conclude that if no beam with a high degree of polarization
is available, it is difficult to observe correlations of higher orders. On the
other hand, from the experimental point of view observing even two-electron
correlations is very difficult \cite{silverman2,kodama}. In comparison to
the~extreme difficulty of, say, a ten-electron correlation
experiment, making a 99\%-polarized electron beam seems to be an easy task
and in this way the spin degree of freedom should have no limitation effect on
the measurement of multi-electron correlations.

\section*{Acknowledgments}
I would like to thank professor M.~Lenc for helpful discussions. This work
was supported by the Czech Ministry of Education, contract No. 144310006.


\begin{thebibliography}{99}
\bibitem{silverman1} M.~Silverman, Il~Nuovo Cimento {\bf 97B}, 200 (1987).
\bibitem{silverman2}M.~Silverman, Phys. Lett.~A {\bf 120}, 442 (1987).
\bibitem{endo} K. Toyoshima and T. Endo, Phys. Lett. A {\bf 152}, 141 (1991).
\bibitem{saito} S.~Saito {\it et~al.}, Phys. Lett. A {\bf 162}, 442 (1992).
\bibitem{ban1} M.~Ban, Phys. Lett. A {\bf 172}, 337 (1993).
\bibitem{tyc} T. Tyc, Phys. Rev. A {\bf 58},  4967 (1998).
\bibitem{europium}E.~Kisker, G.~Baum, A.~H.~Mahan, W.~Raith, and B.~Reihl,
                  Phys. Rev. B {\bf 18}, 2256 (1978).
\bibitem{glauber} R.~Glauber: Quantum theory of coherence, in:
                  Quantum optics, eds. S.~Kay and A.~Maitland, Academic Press,
                  New York 1970.
\bibitem{mandel}L.~Mandel and E.~Wolf: Optical Coherence and Quantum Optics,
                Cambridge University Press, 1995.
\bibitem{kodama} T.~Kodama {\it et al.}, Phys. Rev. A {\bf 57}, 2781 (1998).
\bibitem{silverman3}M.~Silverman, Phys. Lett.~A {\bf 120}, 442 (1987).

\end{thebibliography}
\end{document}